\documentclass[
reprint,
amsmath,amssymb,
aps,
]{revtex4-2}

\usepackage{graphicx}
\usepackage{subfigure}
\usepackage{dcolumn}
\usepackage{bm}
\usepackage{amsmath}
\usepackage{booktabs}
\usepackage{hyperref}

\begin{document}

\preprint{APS/123-QED}
\title{Triaxial Alignment Magnetometer Utilizing Free-Spin Precession in the Geomagnetic Range}
\author{Ge Jin}
\affiliation{School of Instrument Science and Opto-Electronics Engineering, Beihang University, Beijing 100191, China}
\author{Tao Shi}
\affiliation{School of Instrument Science and Opto-Electronics Engineering, Beihang University, Beijing 100191, China}
\author{Sheng Zou}
\email{zousheng@buaa.edu.cn(Corresponding author)}
\affiliation{School of Instrument Science and Opto-Electronics Engineering, Beihang University, Beijing 100191, China}


\begin{abstract}
In this paper, we present a triaxial alignment magnetometer based on free-spin precession deployed in the geomagnetic range. Existing vector measurement methods often require complex optical setups, heating structures, and laser modulation. This study addresses this challenge by employing a linearly polarized probe beam to induce atomic alignment and subsequently detecting the optical polarization rotation caused by the pulsed radio frequency field. The experiment is conducted in a paraffin-coated cell without buffer gas at room temperature, containing rubidium with natural abundance. We report triaxial measurements with a static magnetic field amplitude of approximately 50 $\mu{\text{T}}$ (close to Earth's magnetic field), where the noise levels for each axis are approximately 5.3 ${\text{pT/}}\sqrt{{\text{Hz}}}$, 4.7 ${\text{pT/}}\sqrt{{\text{Hz}}}$, and 9.3 ${\text{pT/}}\sqrt{{\text{Hz}}}$ respectively. The proposed method demonstrates a simple structure suitable for cost-effective and versatile applications.
\end{abstract}

\maketitle



\section{Introduction}\label{sec:Introduction}
Ultrasensitive magnetometers play a crucial role in various research domains and practical applications \cite{budker2007optical}, encompassing medical biomagnetic studies \cite{hamalainen1993MEG,boto2018moving}, aerospace applications \cite{bennett2021precision}, geophysical exploration \cite{renne1988laser,nabighian2005historical}, and fundamental symmetries \cite{kornack2002dynamics,regan2002new}. Most optically-pumped magnetometers (OPMs) operate based on a similar underlying physical principle, relying either on resonant changes in light absorption or rotation of light polarization to determine the magnetic field value. These magnetometer types include Mz \cite{schultze2017optically}, Mx \cite{groeger2006high}, Bell-Bloom \cite{bell1961optically}, spin-exchange relaxation-free (SERF) \cite{ledbetter2008spin}, nonlinear magneto-optical rotation (NMOR) \cite{budker2000sensitive}, and coherent population trapping (CPT) \cite{liang2014simultaneously}, among others. As there are no distinct boundaries separating different methods, numerous researchers have successfully achieved the combination of characteristic features from two or more techniques \cite{grujic2015sensitive,wang2021dual}. The recent investigation on OPMs reveals the potential of compact and ultrasensitive magnetic sensors \cite{shah2013compact,hunter2018free,limes2020portable,tang2021dual,wang2022evaluation,lin2023high}.

Traditionally, OPMs are scalar devices that solely measure the magnitude of the magnetic field while remaining insensitive to its direction. In advanced applications such as navigation and biomedical imaging, it becomes crucial to acquire information about the magnetic field vector. Therefore, modifications have been implemented for vector measurement purposes in most types of magnetometers which typically involve complex setups or additional modulation and demodulation along the magnetic axis or optical axis \cite{seltzer2004unshielded,patton2014all,huang2015three}.
The commonly employed modulation techniques include amplitude- or phase-
modulated optical fields, zero-field resonance \cite{seltzer2004unshielded}, and parametric oscillation \cite{yang2019high}. Among these methods, the SERF-based vector magnetometers exhibit superior performance with femtotesla-level sensitivity. However, their operation requires near-zero magnetic fields and relatively high operating temperatures\cite{lu2022triaxial}. In general, the optical configuration is an orthogonal or parallel pump-probe setup, sometimes requiring an additional repumping laser to increase the effective number of atoms in the probed quantum state \cite{scholtes2016suppression,li2022repumping}. The optical setup poses challenges to miniaturization and array-based measurements.

In this paper, we propose a single-beam optically pumped triaxial magnetometer utilizing free-spin precession (FSP) of atomic alignment at room temperature within the geomagnetic range. The fundamental principle of this scheme involves the creation of atomic alignment and the detection of the FSP signal within a rubidium vapor cell with a natural abundance, coated with paraffin. This cell is subjected to a pulsed radio-frequency (rf) field and illuminated by a linearly polarized probe beam. The presence of a pulsed rf field near the Larmor frequency induces repopulation of atoms within the Zeeman sublevels of the ground state, thereby generating the FSP signal. We determine the triaxial magnetic field by inducing additional magnetic fields onto the triaxial coil and measuring the resulting change in the total magnetic field by FSP process. The utilization of a linearly polarized probe beam and balanced detection, in comparison to the absorption-based scheme, resulted in a significant reduction of common-mode noise and laser intensity fluctuations. The proposed method offers a straightforward single-beam optical setup without the need for heating structures or laser modulation techniques, making it an optimal choice for miniaturization, cost-effectiveness, and versatile applications.

This paper is organized as follows.  First, the density-matrix formalism analysis is elucidated in Section \ref{sec:Principles}. Then, the experimental setup and results are discussed in Sections \ref{sec:Experiment setup} and \ref{sec:Results and discussion}, respectively. Finally, the conclusion is presented in Section \ref{sec:Conclusions}.

\section{Principles}\label{sec:Principles}

\begin{figure}[htbp]
	\centering\includegraphics[width=0.45\textwidth]{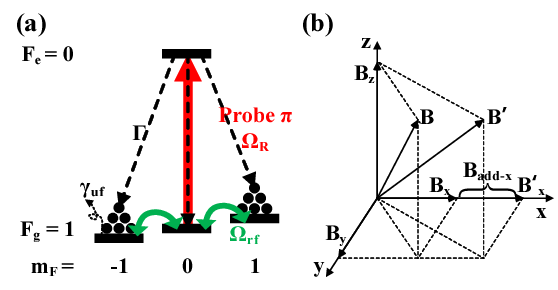}
	\caption{(a) Schematic of simplified $^{85}{\text{Rb}}$ $ \text{D}_1 $ level system. Linearly polarized probe Light (red arrows) produce an aligned ground state. Dashed lines denote the spontaneous transitions at a rate $ \Gamma $. The Zeeman sublevels are coupled by Larmor precession (green curved arrows) driven by a pulsed rf field with frequency $ \omega_{\text{rf}} $ and amplitude corresponding to the rf Rabi frequency ${\Omega _{\text{rf}}}$. (b) The description of the background magnetic field vector ${\mathbf{B}}$ and ${\mathbf{B'}}$ in the laboratory frame.}
	\label{fig:energylevel}
\end{figure}

In our experiment, a single linearly polarized light propagating in the $y$ direction with polarization along the $z$-axis is employed to pump and probe the aligned state of the system. The signals are primarily due to the interaction with the ${F_g} = 3 \to {F_e} = 2$ of the $ \text{D}_1 $ line of $^{85}{\text{Rb}}$, where the subscripts $ g $ and $ e $ indicate the ground and excited states, respectively. For simplicity, we consider the reduction energy-level model of a ${F_g}=1\to{F_e}=0$ transition in Zeeman basis $\left| {{F_g} = 1,{m_{\text{F}}} = 1} \right\rangle $, $\left| {{F_g} = 1,{m_{\text{F}}} = 0} \right\rangle $, $\left| {{F_g} = 1,{m_{\text{F}}} =  - 1} \right\rangle $, and $\left| {{F_e} = 0,{m_{\text{F}}} = 0} \right\rangle $ as shown in Fig. \ref{fig:energylevel}(a). We assume that the static magnetic field $\mathbf{B}$ is dominated along the $z$-axis, corresponding to the Larmor frequency $\Omega_{\text{L}}=\text{g}\mu_{\text{B}}B$ (we set $\hbar  = 1$), where $\mu_{B}$ is the Bohr magneton and g is the Land\'{e} factor. An oscillating magnetic field is applied in the $ x $ direction, $\mathbf{B}_{\mathrm{rf}}=B_{\mathrm{rf}}\cos \omega_{\mathrm{rf}}t \hat{x}$, corresponding to the magnetic Rabi frequency $\Omega_{\mathrm{rf}}=\text{g}\mu_{\text{B}}B_{\mathrm{rf}}$. The Rabi frequency of the optical transition induced by the probe beam is denoted as $\Omega_{R}$. The uniform relaxation of both ground and excited states, including wall collisions, buffer gas collisions, diffusion of atoms from the light beam area, or unpolarized atoms flow from the cell stem, is denoted as $\gamma_{{\rm{uf}}}$. The relaxation of excited states primarily resulting from spontaneous decay is denoted as $\Gamma$.

We describe the experimental process in density-matrix formalism analysis \cite{happer1972optical,budker2002resonant}. The time evolution of density matrix $\rho$ is given by the Liouville equation:
\begin{equation}\label{eq:LiouvilleEquation}
	\frac{{d\rho }}{{dt}} =  - i[H,\rho ] - \frac{1}{2}\{ \xi ,\rho \}  + \Lambda,
\end{equation}
where the $H$ is the total Hamiltonian of the system, $\xi$ and $\Lambda$ are relaxation and repopulation operators, respectively. After making appropriate approximations by considering $\Gamma$ to be significantly larger than all other rates, the general solution can be expressed as \cite{jin2024room,shi2024high}:
\begin{equation}\label{eq:Rho_sol}
	{\rho _{{g_{ \pm 1}}{e_0}}} = \frac{{\left( { - ai + b} \right){{\rm{e}}^{t\left( { - {\gamma _{{\rm{uf}}}} \mp i{\Omega _L}} \right)}}{\Omega _R}}}{{\sqrt 3 \Gamma }},
\end{equation}
with the initial states ${\rho _{{g_{ \pm 1}}{g_0}}} = a + ib$, where $a$ and $b$ are real numbers. From the expectation value of the polarization of the medium, the optical-rotation signal per unit length $dl$ of the medium can be written in the form of:
\begin{equation}\label{eq:phi}
	\frac{{d\alpha }}{{dl}} = \frac{{\sqrt 3 {n_d}\Gamma {\lambda ^2}}}{{2\sqrt 2 \pi {\Omega _R}}}{\rm{Im}}({\rho _{{g_{ - 1}}{e_0}}} - {\rho _{{g_1}{e_0}}}),
\end{equation}
where $\lambda$ refers to wavelength of probe beam, $n_d$ is the atomic density. Combining the equations above, the magnitude of the optical rotation signal can be expressed as:	
\begin{equation}\label{eq:phis}
	\frac{{d\alpha }}{{dl}} = \frac{{b{n_d}{\lambda ^2}}}{{2\sqrt 2 \pi }}{{\rm{e}}^{ - t{\gamma _{{\rm{uf}}}}}}\sin ({\Omega _L}t),
\end{equation}
which resembles an exponential decaying sine-wave with a frequency corresponding to the precession of the rubidium atom, allowing us to measure the total magnetic field through frequency detection.

To measure triaxial magnetic fields, taking the $x$-axis as an example, we introduce an additional magnetic field $B_{\text{add-x}}$ along the $x$-axis using triaxial coils. As depicted in Fig. \ref{fig:energylevel}(b), the initial total magnetic field is denoted as $\mathbf{B}$, and upon application of the supplementary magnetic field, it transforms into $\mathbf{B'}$. Hence, this process can be simplified as follows:
\begin{equation}
	\left\{ \begin{gathered}
		B_x^2 + B_y^2 + B_z^2 = {B^2} \hfill \\
		{({B_x} + {B_{\text{add-x}}})^2} + B_y^2 + B_z^2 = {{B'}^2}.
	\end{gathered}  \right.
\end{equation}
Then the $x$-axis magnetic field can be expressed as:
\begin{equation}\label{eq.Bx}
	{B_x} = ({{B'}^2} - {B^2} - B_{{\text{add-x}}}^2)/(2{B_{{\text{add-x}}}}).
\end{equation}
The magnitude of the total magnetic field $\mathbf{B}$ and $\mathbf{B'}$ can be measured through the FSP process, while the additional magnetic field $B_{\text{add}}$ is determined by the applied current and coil constant of the triaxial coils. By applying $B_{\text{add}}$ in different directions of rectangular coordinate system, we are able to calculate the triaxial magnetic field.

\section{Experiment setup}\label{sec:Experiment setup}

\begin{figure*}[htbp]
	\centering
	\includegraphics[width=1\textwidth]{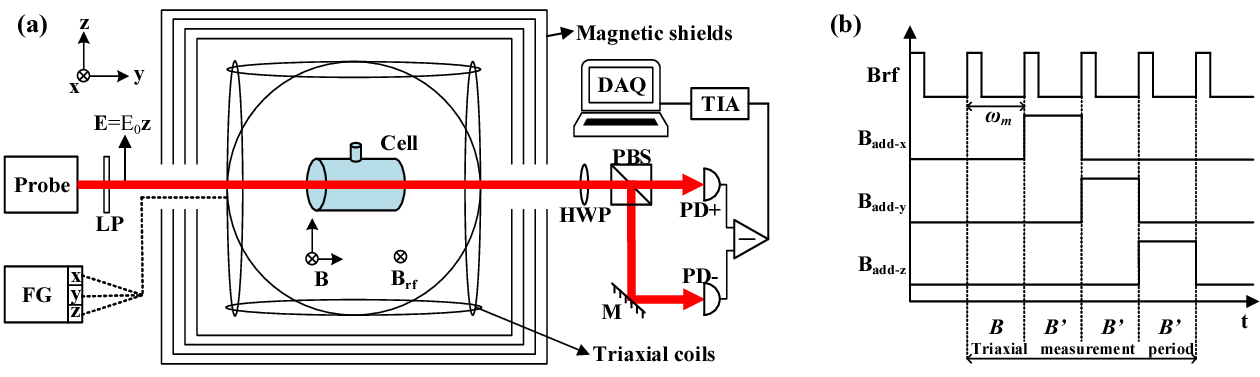}
	\caption{(a) Schematic of the experimental setup. FG, function generator; HWP, half-wave plate; PD, photodiode; PBS, polarization beam splitter; TIA, trans-impedance amplifier; DAQ, data acquisition system. (b) The modulated timing sequence of the rf field $B_{\text{rf}}$ and triaxial extra magnetic field $B_{\text{add-x}}$, $B_{\text{add-y}}$, and $B_{\text{add-z}}$ produced by the triaxial coils (not scaled).}
	\label{fig:schematic of setup}
\end{figure*}

The experimental setup, as illustrated in Fig. \ref{fig:schematic of setup}(a), consists of a cylindrical paraffin-coated cell containing enriched natural abundance rubidium vapor without buffer gas. The dimensions of the cell are 20 mm in diameter and 40 mm in length, while it is enclosed within a four-layer cylindrical $\mu$-metal magnetic shield exhibiting a measured shielding factor of approximately $10^{4}$. The probe beam generated by distributed feedback laser propagates through the cell along the $y$-axis with linear polarization along the $z$-axis. Throughout measurements, the probe beam power remains at about 700 $\mathrm{\mu W}$ with waist radius around 5 $\mathrm{mm}$ and linewidth on the order of 1 MHz. To maximize optical rotation signal, we tune the probe beam on the $^{85}{\text{Rb}}$ $\text{D}_1$ line $F_{g}=3\rightarrow F_{e}=2$ transition.

The three components of the static magnetic field and the pulsed rf field along the $x$-axis are controlled by driving a set of triaxial coils inside the shields using function generators. The rf magnetic field is a pulsed sine wave oscillating near Larmor frequency $ {\Omega _{\text{L}}} $, modulated at frequency of ${\omega _{\text{m}}}$. In Fig. \ref{fig:schematic of setup}(b) shows the modulated timing sequence of the rf field $B_{\text{rf}}$ and triaxial extra magnetic field $B_{\text{add-x}}$, $B_{\text{add-y}}$, and $B_{\text{add-z}}$ produced by the triaxial coils. During each rf pulse period, a frequency related to the total magnetic field can be obtained as the FSP frequency. By continuously adding extra magnetic fields along the $x$-, $y$-, and $z$-axis, triaxial magnetic fields ($B_x$, $B_y$, and $B_z$) can be calculated using Eq. (\ref{eq.Bx}). The sampling rate is set at one-fourth of the entire rf modulating frequency ${\omega _{\text{m}}}$.

After passing through the vapor cell, the polarization of the probe beam is detected using a balanced polarimeter setup consisting of a polarizing beam splitter and two Si PIN photodiodes that quantify the intensities of the two beams emerging from the PBS. The photodiodes have a photosensitivity of approximately 0.6 A/W at 795 nm and a cut-off frequency of 20 MHz. The output differential photocurrent is amplified by a trans-impedance current amplifier and converted into a voltage signal. A band-pass Butterworth filter is applied to eliminate slowly varying terms and high-frequency noise. All data is recorded by the data acquisition system with a sampling rate of 100 MHz and 16-bit sampling accuracy.

\section{Results and discussion}\label{sec:Results and discussion}
\subsection{Single-axis measurement}
\begin{figure}[tbp]
	\centering\includegraphics[width=0.45\textwidth]{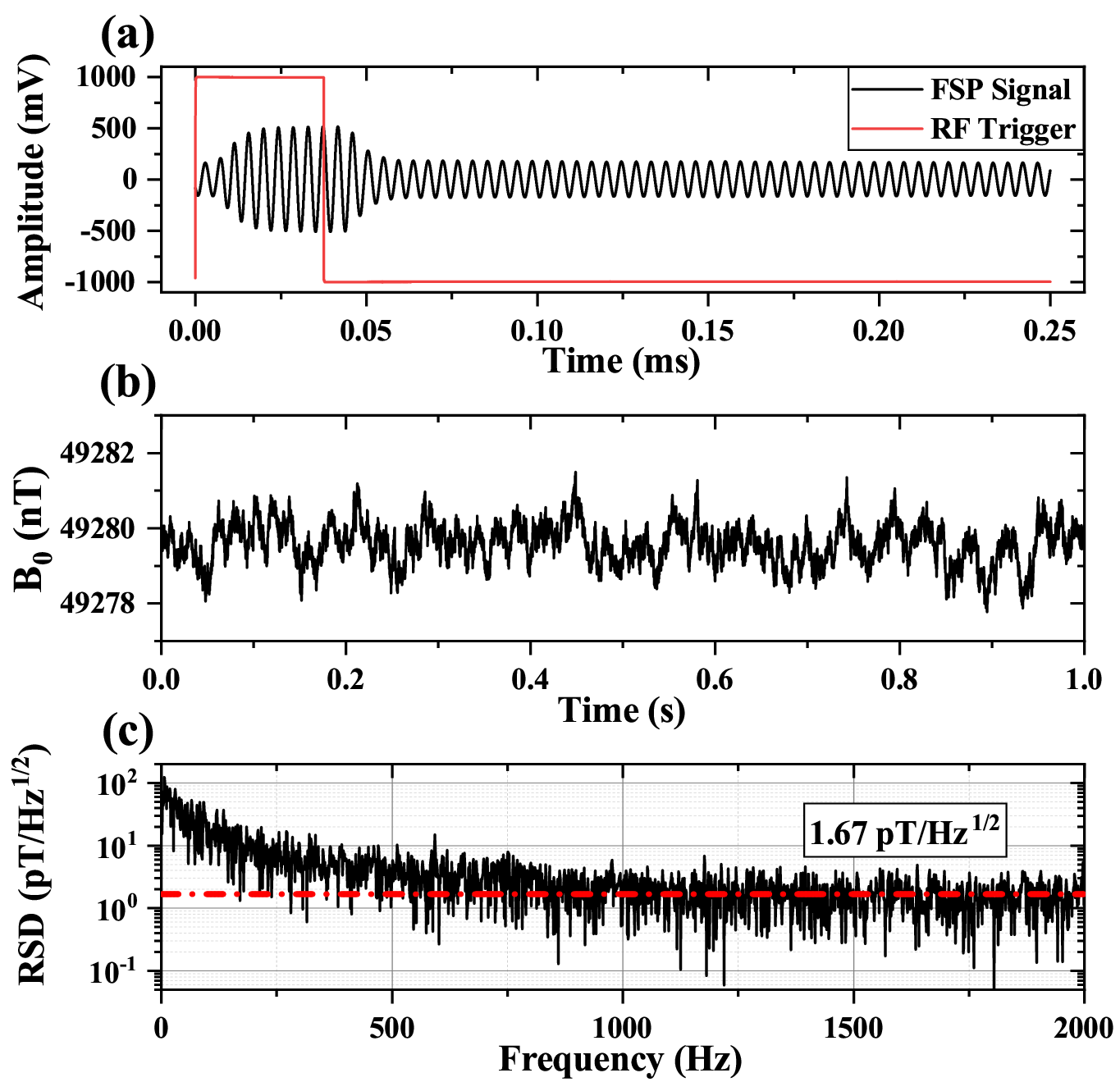}
	\caption{(a) FSP signal (black) in the geophysical magnetic range acquired in a single period with a 4 kHz modulated frequency (red). (b) The magnetic field by fitting the FSP signal over a duration of 1 second. (c) The root spectral density of the time-domain magnetic field data in (b), resulting in a noise floor of 1.67 ${\text{pT}}/\sqrt{{\text{Hz}}}$.}
	\label{fig:FID_B0_RSD}
\end{figure}
First, we evaluate the performance of the single-axis magnetometer in the absence of applied magnetic fields $B_{\text{add}}$. In our experiment, we set the modulation frequency ${\omega _{\text{m}}}$ to 4 kHz and optimized the duty cycle to 15\% in order to enhance the optical rotation signal and facilitate efficient detection of external magnetic fields within a geophysical range of approximately 50 $\mu$T during each period of FSP. The magnetic field we apply is oriented along the $z$-axis and negligible in the orthogonal directions. The rf field frequency is fixed at 230 kHz in close proximity to resonance with the static magnetic field. The change in the magnetic field leads to a slight detuning of the rf field frequency, resulting in a marginal decrease in the amplitude of the output signal. However, it does not affect the FSP frequency. We balance the dc offsets in the raw signals to minimize common-mode noise. Fig. \ref{fig:FID_B0_RSD}(a) shows the trigger signal of the amplitude modulated rf field (red) and a typical FSP signal (black). The observed FSP signal exhibits an approximate delay of 0.01 ms following the RF trigger signal, which we attribute to the time necessary for the atomic population to attain a steady state, thereby limiting the maximum rf field modulation frequency.

To obtain the Larmor frequency and estimate the bias field from the FSP signal train, we employed three methods: the zero-crossing, the fast Fourier transform (FFT), and the fitting algorithm. The zero-crossing algorithm relies on identifying consecutive zero crossing points to determine frequencies. The data acquisition device employed in our study operates at a sampling rate of 100 MHz, whereas the Larmor frequency of $^{85}$Rb in the Earth's magnetic field is approximately 230 kHz. Consequently, each individual Larmor period corresponds to approximately 500 sample points. In order to determine the frequency, we calculate the average time intervals between successive zero crossings during the measurement period. However, it should be noted that minor signal fluctuations occurring around the zero-crossing point can potentially impact the accuracy of this algorithm. The FFT algorithm is utilized to perform a discrete-time Fourier transform on the FSP signal for the extraction of frequency domain characteristic values. In our experiments, we have set the detection frequency at 4 kHz, and each individual detected FSP period consists of 25 k samples. The application of the FFT algorithm to the FSP period results in a spectral frequency interval of 4 kHz, which imposes a fundamental constraint on frequency resolution. Consequently, it becomes challenging to discern subtle variations in magnetic field. The fitting algorithm utilizes the theoretical model derived in Eq. (\ref{eq:phis}) to fit the FSP signal with a relatively slow processing speed but high accuracy, making it suitable for both low- and high-bias magnetic field situations. We enhance the fitting speed by pre-fitting to determine the initial value of the fitting function. The data stream is recorded over a duration of 1 s and subsequently subjected to processing by fitting each individual FSP trace. The extracted consecutive magnetic field data are shown in Fig. \ref{fig:FID_B0_RSD}(b). 

Noise spectral density serves as a meaningful indicator of magnetometer performance. After performing a discrete-time Fourier transform, the root spectral density (RSD) of the measured magnetic field is illustrated in Fig. \ref{fig:FID_B0_RSD}(c). It can be observed that the noise floor of the magnetometer is approximately 1.67 ${\text{pT}}/\sqrt{{\text{Hz}}}$. The bandwidth of the magnetometer is mainly limited by the modulating frequency ${\omega _{\text{m}}}$ in accordance with the Nyquist theorem.

\subsection{Triaxial measurement}
\begin{figure}[thbp]
	\centering\includegraphics[width=0.45\textwidth]{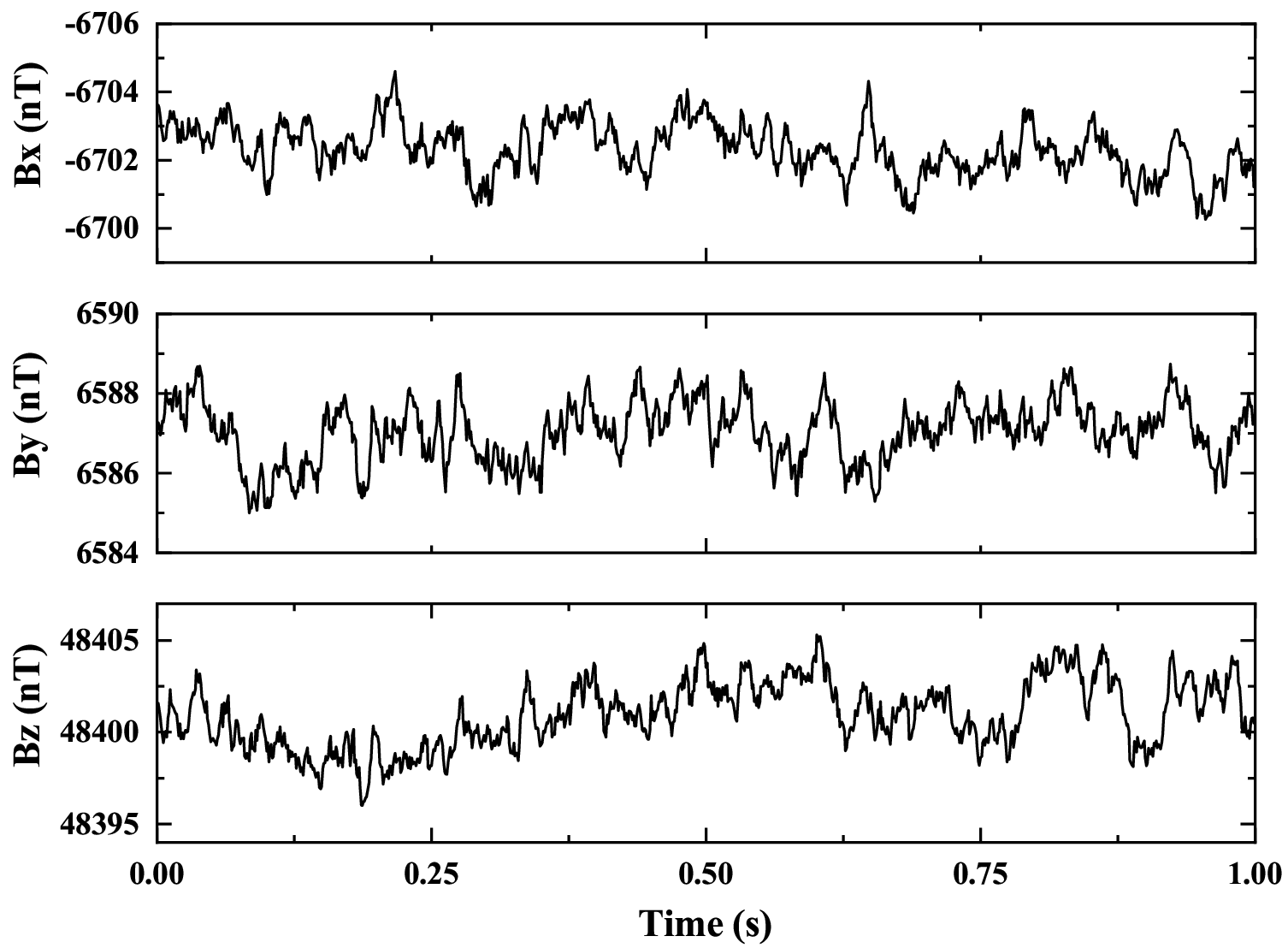}
	\caption{The triaxial magnetic field components ${B_x}$, ${B_y}$, and ${B_z}$ recorded over a duration of 1 second.}
	\label{fig:triaxialB}
\end{figure}

\begin{figure}[thbp]
	\centering\includegraphics[width=0.45\textwidth]{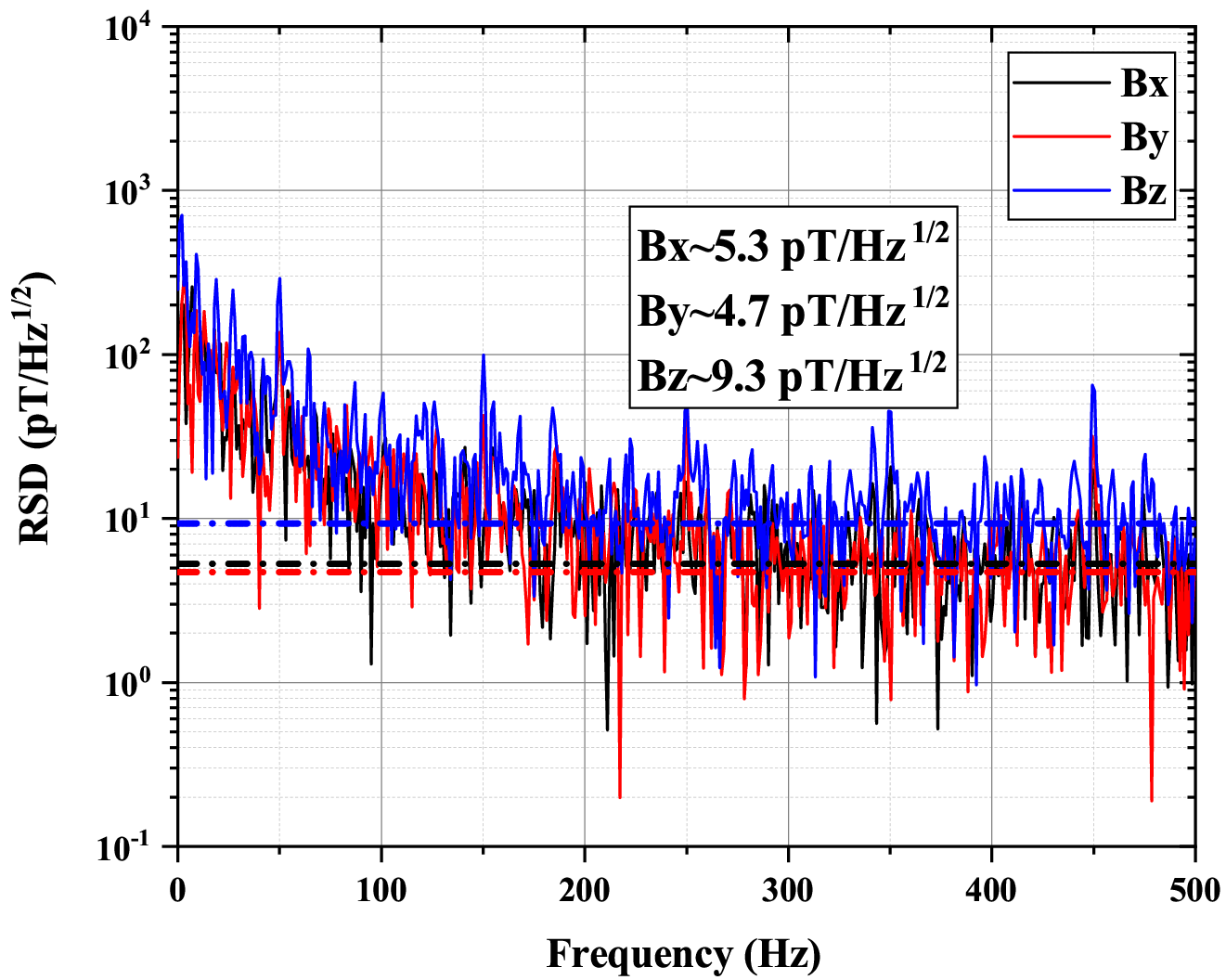}
	\caption{The root spectral density of the time-domain triaxial magnetic field. The noise floor are approximately 5.3 ${\text{pT/}}\sqrt{{\text{Hz}}}$, 4.7 ${\text{pT/}}\sqrt{{\text{Hz}}}$, 9.3 ${\text{pT/}}\sqrt{{\text{Hz}}}$ for $x$-, $y$-, and $z$-axis, respectively.}
	\label{fig:triaxialRSD}
\end{figure}

For triaxial magnetic field measurements, we introduce a bias magnetic field along the $z$-axis while also incorporating components in the $x$ and $y$ directions to simulate geophysical conditions. 
The triaxial coil constants and crosstalk are calibrated using a flux-gate magnetometer, as illustrated in table \ref{tab:triaxialcoil}. Through the FSP measurement process, we obtain the magnetic field values before and after the introduction of an additional magnetic field ${B_{\text{add}}}$, denoted as $B$ and $B'$ respectively. Subsequently, triaxial extra magnetic fields are applied sequentially. The resulting triaxial magnetic field is then calculated at a sampling rate of 1 kHz, as depicted in Fig. \ref{fig:triaxialB}. 

\begin{table}[thbp]
	\centering
	\caption{Triaxial coil constant (nT/V) \& Crosstalk}
	\begin{tabular}{
			>{\centering\arraybackslash}p{1.8cm}
			>{\centering\arraybackslash}p{1.8cm}
			>{\centering\arraybackslash}p{1.8cm}
			>{\centering\arraybackslash}p{1.8cm}}
		\toprule
		Axis       & X          & Y          & Z \\
		\midrule
		${B_{\text{add-x}}}$          & -6547.96   & 2\%        & 1.06\% \\
		${B_{\text{add-y}}}$          & -1.81\%    & 6525.21    & -0.52\% \\
		${B_{\text{add-z}}}$          & -1.12\%    & 0.19\%     & 4881.23 \\
		\bottomrule
	\end{tabular}
	\label{tab:triaxialcoil}
\end{table}

\begin{figure}[thbp]
	\centering\includegraphics[width=0.45\textwidth]{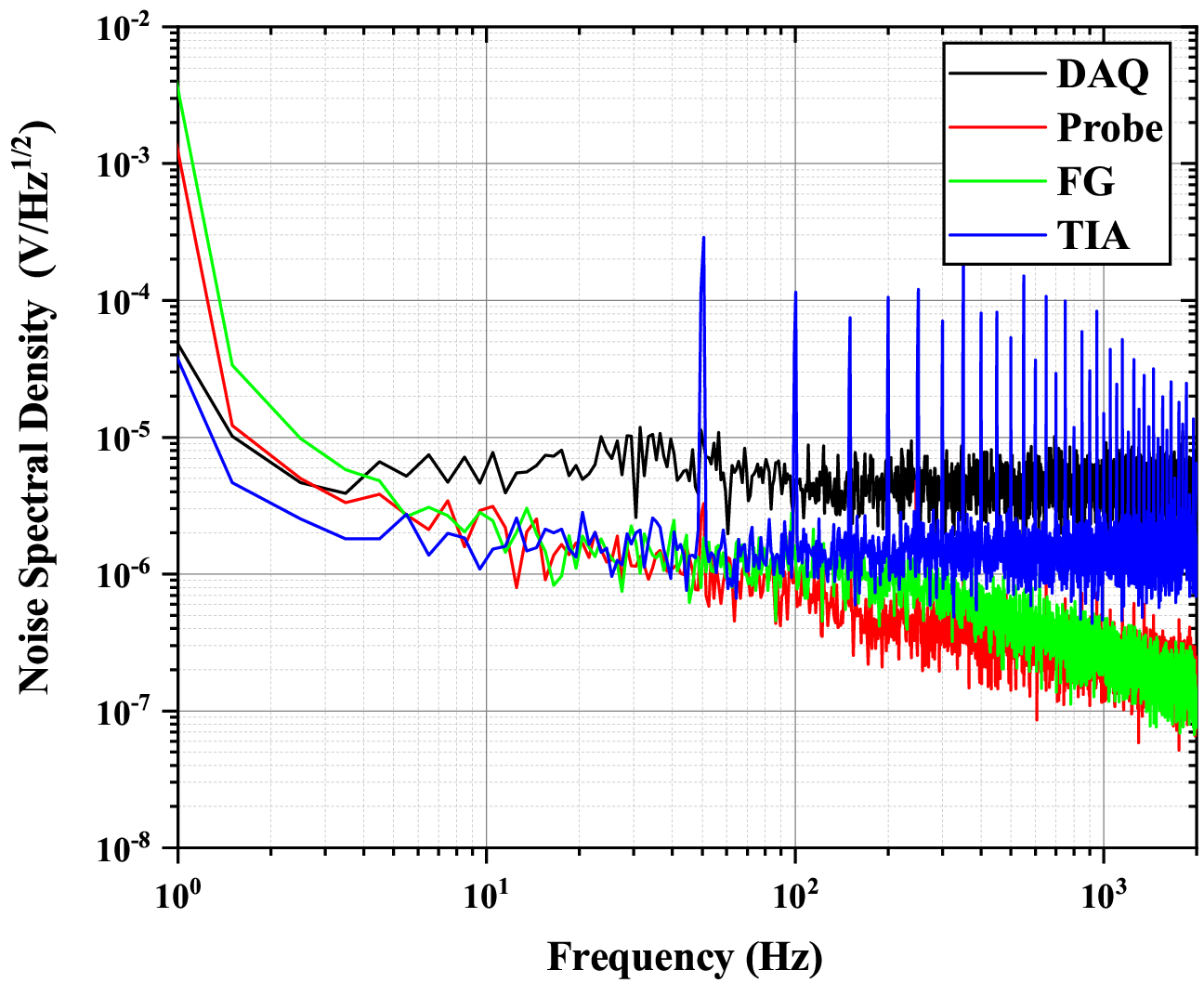}
	\caption{Noise spectral density of the data acquisition system, probe intensity, function generator, and trans-impedance amplifier.}
	\label{fig:noise}
\end{figure}

The noise floor of each axis in Fig. \ref{fig:triaxialRSD} is determined by performing a discrete time Fourier transform on the acquired data, yielding values of approximately 5.3 ${\text{pT/}}\sqrt{{\text{Hz}}}$, 4.7 ${\text{pT/}}\sqrt{{\text{Hz}}}$, and 9.3 ${\text{pT/}}\sqrt{{\text{Hz}}}$ for the $x$-, $y$-, and $z$-axis, respectively. The noise floor of the triaxial measurement is evidently higher compared to that of the single-axis measurement, primarily due to the crosstalk coupling among the triaxial coils and the interdependence between two non-independent measurements of magnetic field $B$ and $B'$. The bandwidth of the triaxial magnetometer is reduced to one quarter of its initial value due to the inclusion of four measurements within each triaxial measurement period. The modulation frequency should be appropriately limited to avoid excessive transient effects of the atoms \cite{jin2019transient}. The rising and falling edge of the extra magnetic fields is about 10 ns in the experiment. In our previous work, we use the hyperfine-selective optical pumping of ${F_g} = 2$ to ${F_e} = 2$ populates the atoms into Zeeman sublevels of ${F_g} = 3$, thereby increasing the population of sensitive atoms. This approach demonstrates a tenfold increase in the amplitude of the optical rotation signal and exhibits an enhanced sensitivity that is nearly four times greater than that of the no-repump configuration. However, this proposed method necessitates an additional laser as a repump source, which increases system complexity and poses challenges for integration. Consequently, we prioritize performance by simplifying the optical setup and eliminating the need for heating structures and laser modulation, thereby making it an ideal choice for cost-effective and versatile applications.

The noise spectral density of the data acquisition system, probe intensity, function generator and trans-impedance amplifier is illustrated in Fig. \ref{fig:noise}. The noise density of the DAQ under no-input conditions is approximately 4.5 $\mu {{\text{V}}_{{\text{rms}}}}/\sqrt {{\text{Hz}}}$, surpassing other noise sources by one or two orders of magnitude. According to the theory of Cram$\acute{\text{e}}$r-Rao lower bound, the derived noise density depends on both the signal-to-noise ratio and the number of data points \cite{yao1995cramer}. We prioritize the high sampling rate and resolution of the acquisition system, making necessary trade-offs in terms of noise.

\section{Conclusions}\label{sec:Conclusions}
In summary, we present an FSP alignment magnetometer subjected to a pulsed RF field, achieving triaxial sensitivities of 5.3 ${\text{pT/}}\sqrt{{\text{Hz}}}$, 4.7 ${\text{pT/}}\sqrt{{\text{Hz}}}$, and 9.3 ${\text{pT/}}\sqrt{{\text{Hz}}}$ in the Earth's magnetic field along the $x$-, $y$-, and $z$-axis, respectively. The use of a linearly polarized probe beam combined with balanced detection results in a significant reduction in common-mode noise compared to absorption-based detection methods. This approach is implemented in a straightforward single-beam configuration without the need for heating structures or laser modulation techniques, thus conferring a notable advantage in sensor miniaturization. A limitation of this approach is the simultaneous and independent nature of the triaxial measurements, leading to an increase in noise levels. Additionally, the presence of imperfect orthogonality compromises the accuracy of the magnetometer in measuring both the direction and magnitude of the applied field. In future research, the integration of an electronic control system and optimized coil design will significantly enhance the practical applicability of this approach.

\section{Acknowledgments}\label{sec:Acknowledgments}
This work is supported by National Natural Science Foundation of China (NSFC) No. 52205549, Beijing Natural Science Foundation NO. 1222025, Central Guided Local Science and Technology Development Funding Program NO. 2023ZY1005, and Foundation from Science and Technology on Inertial Laboratory No. 2022-JCJQ-LB-070-02.
\nocite{*}
\bibliography{myref}
\end{document}